\documentclass[prb,twocolumn,aps]{revtex4-2}
\usepackage{bm}
\usepackage{amsmath}
\usepackage{amssymb}
\usepackage{times}
\usepackage[usenames,dvipsnames]{xcolor}
\usepackage{graphicx}
\usepackage{dcolumn}
\usepackage{simplewick}
\usepackage{hyperref}

\hypersetup{
  colorlinks=false,
  colorlinks=true,
  citecolor=blue,
  linkcolor=blue,
  urlcolor=blue}

\begin{document}

\title{Intralayer antiferromagnetism in two-dimensional van der Waals magnet Fe$_3$GeTe$_2$}

% Force line breaks with \\

\author{Neesha Yadav}%
\affiliation{Department of Physics, Indian Institute of Technology, Delhi, 
Hauz Khas, New Delhi, India - 110016}%Lines break automatically or can be forced with \\

\author{Shivani Kumawat}%
\affiliation{Department of Physics, Indian Institute of Technology, Delhi, Hauz Khas, New Delhi, India - 110016}%

\author{Sandeep}%
\affiliation{Department of Physics, Indian Institute of Technology, Delhi, Hauz Khas, New Delhi, India - 110016}%
%\email{pintu@physics.iitd.ac.in}
%\affiliation{ 
%Department of Physics, Indian Institute of Technology, Delhi, New Delhi, India %\\This line break forced with \textbackslash\textbackslash
%}%
\author{Brajesh Kumar Mani}
\affiliation{Department of Physics, Indian Institute of Technology, Delhi, Hauz Khas, New Delhi, India - 110016}%
\author{Pintu Das$^*$}
 \affiliation{Department of Physics, Indian Institute of Technology, Delhi, Hauz Khas, New Delhi, India - 110016}%%\homepage{http://www.Second.institution.edu/~Charlie.Author.}
%\affiliation{%
%Second institution and/or address%\\This line break forced% with \\
%}%

\date{\today}% It is always \today, today,
             %  but any date may be explicitly specified

%%%%%%%%%%%%%%%%%%%%%
\begin{abstract}
%%%%%%%%%%%%%%%%%%%%%

For the van der Waals magnet Fe$_3$GeTe$_2$, although a ferromagnetic ground state has been reported, there are also reports of complex magnetic behavior suggesting coexistence of ferromagnetism and antiferromagnetism due to the intricate interaction between Fe$^{+3}$ and Fe$^{+2}$ ions in this system. The exact nature of the interactions and the origin of antiferromagnetism are still under debate. Here, we report the observation of signature of ferromagnetic and antiferromagnetic couplings between different Fe-ions in the anomalous Hall effect measured for devices of mechanically exfoliated Fe$_3$GeTe$_2$ nano-flakes of thicknesses ranging from\,$\sim$\,15-20 layers. The temperature-dependent anomalous Hall effect data reveal two sharp step-like switchings at low temperature ($T\lesssim150\,$K). Our detailed analyses suggest the step-like sharp switchings in anomalous Hall resistance are due to the magnetization reversal behavior of different Fe-ions in individual layers of Fe$_3$GeTe$_2$. The experimental results can be explained by considering an intra-layer antiferromagnetic coupling between Fe$^{+3}$ and Fe$^{+3}$ ions, whereas intra-layer ferromagnetic coupling between Fe$^{+3}$ and Fe$^{+2}$ in the system. Our experimental results and the analyses are supported by the first-principles calculations for energetics and intralayer as well as interlayer exchange coupling constants.

\end{abstract}

\date{\today}

\maketitle

%%%%%%%%%%%%%%%%%%%%%%%%%%%%%%%
\section{Introduction}
%%%%%%%%%%%%%%%%%%%%%%%%%%%%%%%

The discovery of long-range magnetic order in two-dimensional van der 
Waals (vdW) magnets has opened up possibilities to explore fundamental 
questions, as well as for applications in future energy-efficient 
miniaturized spintronic devices. A strong anisotropy, which is a 
prerequisite for long-range order in lower dimensions, leads to 
the opening of an energy gap in the spin wave excitation as well 
as suppression of thermal fluctuations. Two-dimensional magnets, 
such as transition metal (TM) halides, TM-chalcogenides, and Fe$_3$GeTe$_2$ 
(FGT-3), exhibit a multitude of layered-dependent properties.
However, due to the stability of FGT-3, etc., in air and the 
possibility of raising the $T_{\rm{C}}$ to about room temperature 
\cite{deng2018gate}, this class of materials has attracted significant 
attention from researchers in recent years. FGT-3 shows itinerant 
ferromagnetic behavior due to Fe$^{+3}$ ions and localized behavior 
due to Fe$^{+2}$ ions \cite{sharma2023non}. The itinerant nature 
leading to the interplay of charge and spin degrees of freedom
in the system makes it an excellent candidate for applications in 
spintronics. New properties are being explored by the formation 
of heterostructures using FGT-3 with other relevant materials. 
Several groups have reported interesting results of chiral spin 
textures in heterostructures formed with FGT-3 and spin-orbit-coupled 
materials such as WTe$_2$, graphene \cite{wu2020neel, srivastava2024unusual}. 
Furthermore, the material shows Kondo scattering behavior at low temperatures. 
Even more interestingly, a direct correlation was found between the 
ferromagnetism and the effective electron mass, suggesting the system to 
exhibit the behavior of heavy fermions in this 3d itinerant
magnetic system with partially filled d-band \cite{zhang2018emergence}.

%%%%%%%%%%%%%%% Fig.1
\begin{figure}[h!]
\includegraphics[width=1\linewidth]{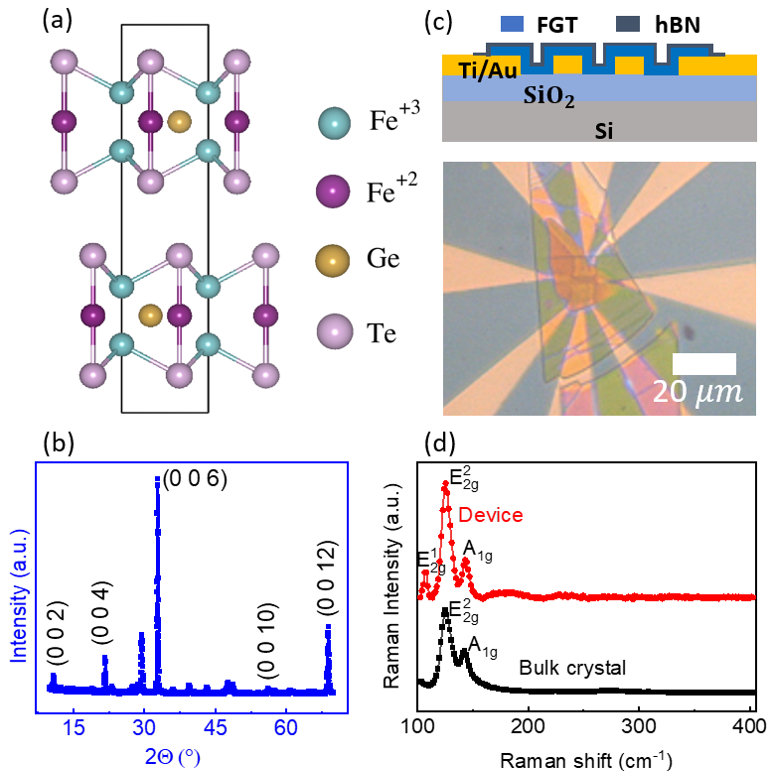}
\caption{Crystal structure and device fabrication of FGT-3. 
	(a) The crystal structure of a bilayer FGT-3. Monolayer FGT-3 contains 
	two types of Fe atoms, Fe$^{+3}$ and Fe$^{+2}$, at different sites. (b) 
	The X-ray diffraction pattern of FGT-3 single crystal. (c) Schematic 
	diagram of the device geometry (top panel). The optical micrograph of 
	fabricated FGT-3/hBN device (bottom panel). (d) Raman spectra of 
	FGT-3 single crystal (black) \& the nanoflake of FGT-3 device (red).}
\label{Structural properties}
\end{figure}
%%%%%%%%%%%%%%%%%%%%

Although the system clearly exhibits ferromagnetic behavior in the entire
ordered range \cite{deng2018gate}, some anomalies in the magnetization versus
temperature ($M-T$) as well as $T$-dependent local magnetic imaging of domains
using magnetic force microscopy measured on bulk crystals grown by chemical
vapor transport (CVT) method lead to the suggestion of coexisting ferromagnetic
and complex antiferromagnetic phases in a wide temperature range with an
antiferromagnetic ground state \cite{yi2016competing}. Later, a kink in the
$T$-dependent Hall response of mechanically exfoliated thin layers of FGT-3
(thickness\,$\ge30$\,nm) was suggested to be a signature of competing
ferromagnetism and interlayer antiferromagnetism in the
system \cite{chyczewski2023probing}. The exfoliation was performed in a glove
box with a significant oxygen environment of about 20 ppm, and thus, there is
an uncertainty of the quality of the devices. These results suggested an
interlayer antiferromagnetic coupling between the ferromagnetic Fe-atom planes.
However, a few other works suggested the antiferromagnetic phase exhibiting the
exchange bias effect is due to the oxidized FGT-3
layer~\cite{kim2019antiferromagnetic}. A recent work based on analysis of
ac-susceptibility results suggested 2D spin-glass behavior in the FGT-3
system~\cite{pal2024realization}. Thus, it is clear that inspite of a large
number of investigations carried out for this class of systems, a clear
understanding of the magnetic ordering of the systems has not been achieved so
far. A detailed analysis of magnetization as well as magnetotransport behavior
of the same crystals may be helpful to converge the actual physics governing
the antiferromagnetic behavior of the system. An understanding of this aspect
is critical for the use of the material in any real spintronic devices.

In this work, we have carried out a detailed analysis of the magnetization 
of bulk crystals and the magneto-transport behavior of thin film of 
FGT-3 samples exfoliated in a very inert atmosphere. To avoid ambiguity 
and sample-to-sample variations,the same crystals were used for both 
investigations. For magneto-transport behavior, primarily the anomalous 
Hall effect was probed so that a direct correlation between the magnetization 
and the Hall effect is established. Together with the detailed first-principles
calculations, our experimental results provide a clear indication of 
an intra-layer antiferromagnetic coupling, in addition to a strong 
interlayer ferromagnetic coupling, which so far has been undetected.

%%%%%%%%%%%%%%%%%%%%%%%%%%%%%%%%%%%%%%%%%%%%%%%
\section{Experimental and Computational Details}
%%%%%%%%%%%%%%%%%%%%%%%%%%%%%%%%%%%%%%%%%%%%%%%

Single crystals of FGT-3 were grown by CVT method and were obtained from
commercial sources (viz., HQ graphene). The crystals were characterized by
X-ray diffraction and Raman spectroscopy experiments performed at room
temperature. Magnetic measurements were carried out using a Magnetic Property
Measurement System (make: Quantum Design) under zero field cool (ZFC) and Field
cooled cool (FCC) protocols. For the analysis of thermal hysteresis,
measurements of magnetization ($M$) were performed using field cooled warming
(FCW) protocol.

For measurements of anomalous Hall effect (AHE), mechanically exfoliated thin
layers of FGT-3 devices encapsulated with hexagonal boron nitride (hBN) were
used. The same batch of single crystals, as used for the magnetic measurements,
was chosen for the transport measurements. The flakes were exfoliated by the
conventional scotch tape method on the polydimethylsiloxane (PDMS) and then dry
transferred on Hall cross geometry pre-patterned on Si/SiO$_2$ substrates by
optical lithography followed by electron-beam deposition of Ti/Au and lift-off.
Further, the transfer process was repeated for hBN without breaking the vacuum
to prevent oxidation of the topmost FGT-3 layer. The entire fabrication process
was carried out in an inert atmosphere, inside an argon-filled glovebox with
O$_2$ as well as H$_2$O level of  $\leq$ 0.1\,ppm. The bottom contact was
chosen so that the entire process of transfer and encapsulation can be carried
out without breaking the vacuum. The devices were wire bonded to leadless chip
carriers for measurements in a Physical Property Measurement System (PPMS)
(make: Cryogenics, UK).

%%%%%%%%%%%%%%%%%%%%% fig2
\begin{figure}
\includegraphics[width=1\linewidth]{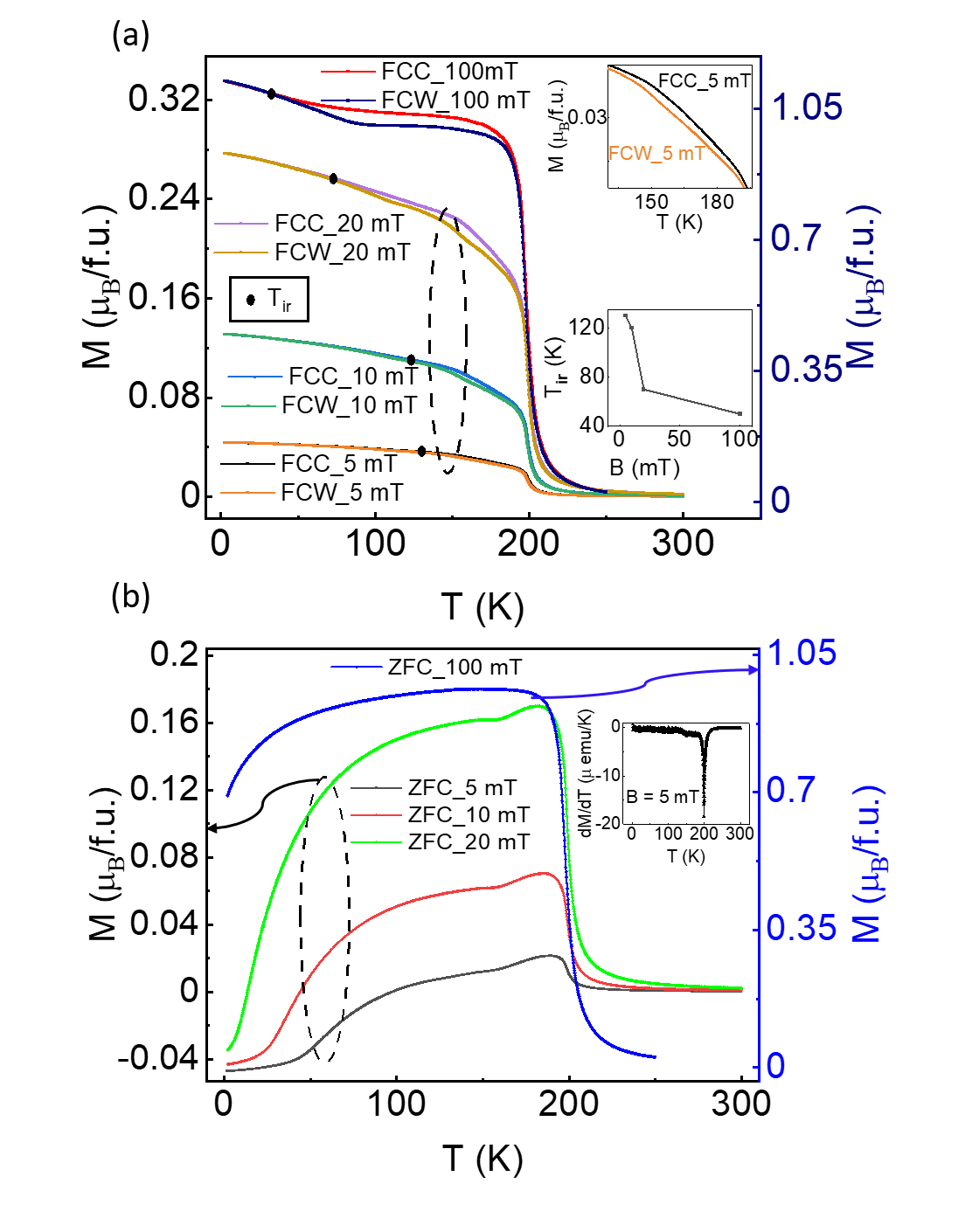}
\caption{(a) The magnetization at different 
	applied fields ranging from 5\,mT to 100\,mT showing the thermo-hysteresis 
	between FCC \& FCW curves. The top and bottom right insets indicate the 
	zoomed $M$-$T$ curve at the field of 5\,mT and the irreversibility temperature 
	($T_\mathrm{ir}$) at different applied fields, respectively. (b) ZFC curves at 
	various applied magnetic fields showing a kink near 160\,K. Inset shows the 
	derivative of magnetization at 5\,mT, indicating the $T_\mathrm{C}$ = 198.6\,K.}
\label{Magnetization}	
\end{figure}
%%%%%%%%%%%%%%%%%%%%%%%%%%%

To support our experimental results, we performed density functional theory
(DFT) based first-principles calculations using the projected augmented wave
(PAW) method \cite{blochl1994projector} as implemented in the {\it Vienna Ab
initio Simulation Package} (VASP)
\cite{kresse1999ultrasoft,kresse1994ab,kresse1996efficient}. The
exchange-correlations among electrons were incorporated using the generalized
gradient approximation (GGA) based Perdew-Burke-Ernzerhof (PBE) pseudopotential
\cite{perdew1996generalized}. All calculations employed a plane-wave basis with
an energy cutoff of 500 eV. The energy and force convergence criteria were set
at $10^{-6}$ and $10^{-5}$ eV, respectively, for all self-consistent-field
calculations. The conjugate gradient algorithm was employed for structural
relaxation, performed on a $\Gamma$-centered k-grid of 16$\times$16$\times$4.
To incorporate the effect of strongly correlated $3d$-electrons of Fe, on-site
Coulomb interaction was included using the rotationally invariant DFT+U
approach of Dudarev et al. \cite{cococcioni2005linear}. Our computed U value, 4.4
eV, is in good agreement with the reported value 4.3 eV and 4.8 eV \cite{ghosh2023unraveling}. However, it
is reported that the DFT+U method cannot give an accurate magnetic moment of Fe
atoms \cite{zhuang2016strong,ghosh2023unraveling}. It overestimates the
magnetic moment by almost 1 $\mu_B$. Therefore, for magnetic moment
calculation, we ignore U correction.

%%%%%%%%%%%%%%%%%%%%%%%%%%%%%%%%%%%%%%%%%%%%%%%%%%
\section{Results \& Discussion}
%%%%%%%%%%%%%%%%%%%%%%%%%%%%%%%%%%%%%%%%%%%%%%%%%%%

Fe$_3$GeTe$_2$ crystallizes in space group of $P6_3/mmc$ has a layered
structure with Fe$_3$Ge layer sandwiched between two Te layers as shown in
Fig.\,\ref{Structural properties}(a). The Fe$_3$Ge layer consists of two
inequivalent Fe atoms, viz., Fe$^{+3}$ and Fe$^{+2}$. Each monolayer of FGT-3
is $\sim$\,0.8\,nm thick. The unit cell structure consisting of bilayers with
van der Waals gap of 0.295\,nm between two adjacent layers is shown in
Fig.\,\ref{Structural properties}(a).

The crystal structure and phase purity of the crystals are confirmed from
powder-X-ray diffraction (XRD) measurements performed at room temperature using
an X-ray diffractometer (make: Malvern PANalytical; MODEL: Empyrean) with
Cu-K$\alpha$ radiation as source. The XRD pattern shows in the prominent
(0,0,2n) peaks where n = 1, 2, 3, ..., A few additional peaks from other
different planes from the crystallites in the powder, typical for a powder 
XRD pattern from, are also observed (Fig.\,\ref{Structural
properties}(b))~\cite{bera2023enhanced}. The XRD pattern confirms the phase
purity of the crystals. The Raman spectroscopy result for the single crystal
(bulk) FGT-3 is shown in Fig.\,\ref{Structural properties}(d). The spectrum
shows peaks at 125\,cm$^{-1}$  \& 143\,cm$^{-1}$ corresponding to
$E_\mathrm{2g}^2$  and $A_\mathrm{1g}$ modes respectievly. A hump-like feature
at 103\,cm$^{-1}$ is observed, which is suggested to be the position of
$E_\mathrm{2g}^1$ peak with significantly reduced intensity in bulk crystal
~\cite{kong2021thickness}. The observed modes are consistent with the reported
results, further demonstrating the purity of the crystals.

Fig.\,\ref{Magnetization} shows detailed magnetization data measured under ZFC,
FCW, and FCC protocols. $T_{\rm{C}}$, as determined from the derivative of
$M-T$ data measured at 5\,mT field applied out of plane (easy axis) of the
sample, is found to be 198.6\,K (see inset of Fig.\,\ref{Magnetization}(b)).
ZFC and FCW measurements show a bifurcation in the temperature range of 30\,K
to 130\,K depending on the applied field. Such splitting of ZFC and FCW data
suggests magnetic irreversibility, typically observed in spin-glass systems.
However, this behavior alone is not a direct test of spin-glass behavior and
may indicate other complex phenomena (see discussions below). Further, a kink
is observed at a temperature close to 160\,K- in low-field ZFC curves which
vanishes in the higher field data (e.g., 100\,mT), see Fig.\,\ref{Magnetization}
(b). The kink is further investigated by performing FCC and FCW measurements. 
A hysteresis is observed in the FCC and FCW data at
around the same temperature where a kink is observed in ZFC data (see
Fig.\,\ref{Magnetization}(b)). This indicates that the magnetic moments 
do not follow the same orientation dynamics with temperature during 
cooling and warming cycles. Interestingly, the position of the 
hysteresis monotonically shifts to lower $T$ as the field is increased 
from 5\,mT to 100\,mT. This is clearly observed from the irreversibility 
temperature ($T_\mathrm{ir}$), marked by closed circles in the 
corresponding $M-T$ plots.  $T_\mathrm{ir}$ refers to the temperature 
at which the bifurcation between FCC and FCW starts while
warming up. The lower inset in Fig.\,\ref{Magnetization}a shows the plot of
$T_\mathrm{ir}$ vs $T$. The shift in $T_\mathrm{ir}$ with $T$ is tabulated in table \ref{tab:table1}. Moreover, at low temperature, the ZFC curve measured at
a small OOP field of up to 20 mT reduces to a negligible value of
magnetization. The behavior of thermo-hysteresis in magnetization suggests the
presence of different magnetic interactions, such as due to the coexistence of
magnetic phases, in the system~\cite{a53}.

%%%%%%%%%%%%%%%%% fig 
\begin{figure}
\includegraphics[width=1\linewidth]{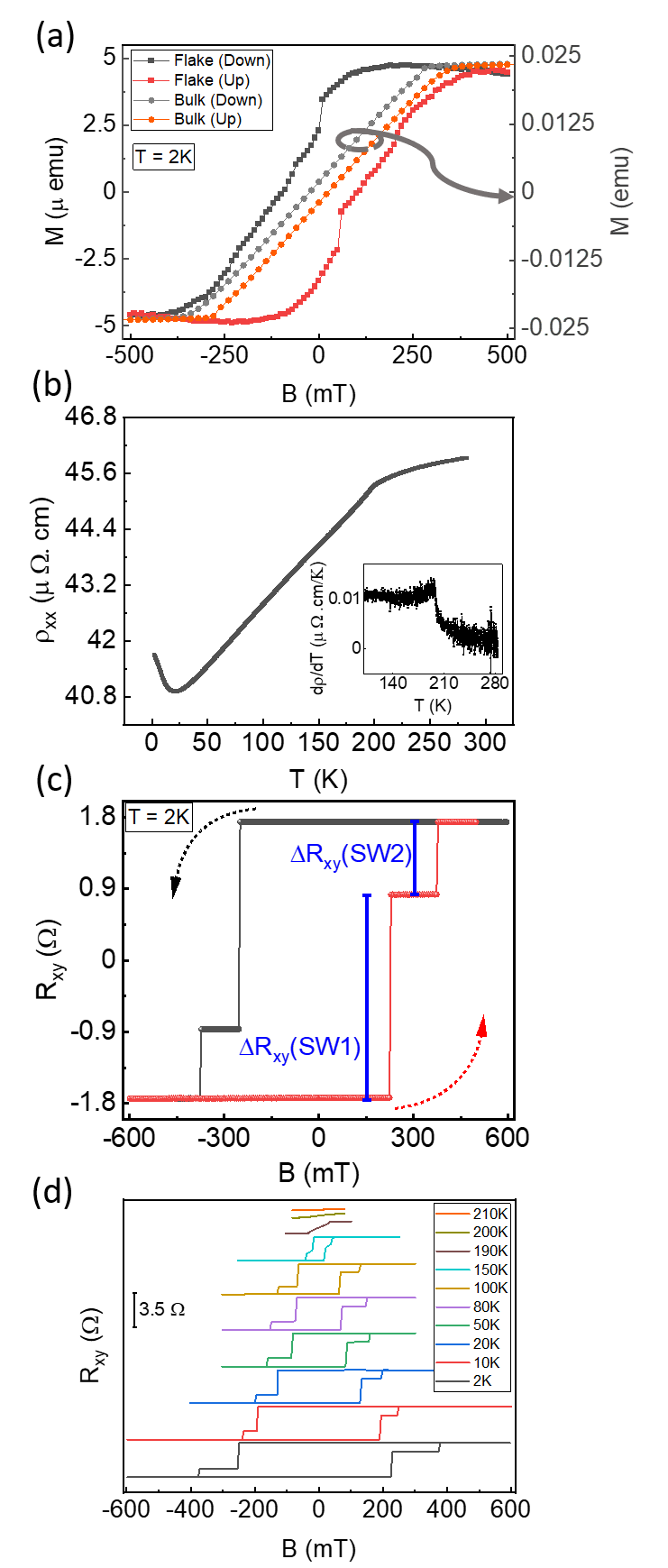}
\caption{The magneto-transport behavior of FGT-3/hBN device.
	(a) \textit{M-H} of few-layer flakes and bulk single crystal of FGT-3 at 
	\textit{T}\,=\,2\,K. (b) The temperature-dependent resistivity of a few-layer 
	($\sim$\,20 layers) FGT-3 device. (Inset) The derivative of the resistivity 
	curve showing the $T_\mathrm{C}$ of 196\,K. (c) Anomalous Hall effect of a 
	$\sim$\,20 layers FGT-3 device at \textit{T}\,=\,2\,K. (d) The temperature 
	evolution of the anomalous Hall behavior of the FGT-3 device.}
\label{Transport}
\end{figure}
%%%%%%%%%%%%%%%%%%%%%%%

Earlier, Yi, \textit{et al.} had reported observation of such thermohysteresis
and kinks in corresponding $M-T$ results obtained from CVT-grown FGT crystals.
Thus, our results of the magnetization of bulk crystals are very consistent
with those reported for the CVT-grown sample. The unusual behavior of
magnetization under different conditions was suggested by Yi, \textit{et al.}
to be due to the presence of competing ferrro-(FM) and antiferro-magnetically
(AFM) ordered Fe moments in the system~\cite{yi2016competing}. Based on the
magnetization and ac-susceptibility measurements, an AFM ground state as well
as a coexisting AFM and FM states in the temperature range of
$\sim$150\,-\,200\,K was suggested to be responsible for the thermohysteresis
behavior. Additionally, recently Chyczewski \textit{et al.} have reported an
unusual behavior in magnetotransport characteristics of exfoliated Fe-doped FGT
crystals~\cite{chyczewski2023probing}. Peaks in the $dR_{xy}/d(\mu_0H)$ as well
as MR vs $T$ as observed at $T\sim120$\,K were considered as the signature for
an AFM phase in the crystal. Both of these works suggested an interlayer AFM
coupling between Fe ions, coexisting with in-plane ferromagnetic interactions
between other Fe ions, for this system. We note here that although an AFM spin
ordering was suggested as the origin of an unusual peak observed in the
magnetotransport data, no direct evidence of this has been found in the
transport data so far. Moreover, the literature, which is very limited on this
problem, shows contradictory suggestions on the AFM interactions in the
system~\cite{ke2020magnetic}. Thus, it is clear that a detailed investigation
of this aspect is absent in the literature.

The analysis of magnetization switching behavior of individual layers of 
FGT-3 crystals may provide important insight into this problem.
Fig.\,\ref{Transport}(a) shows the magnetization behavior of few-layered flakes
along with that of a bulk FGT-3 crystal measured by SQUID magnetometry in an
MPMS. Contrary to the continuous change in magnetization observed for bulk
crystals, interestingly, a sharp jump, along with a few irregularities at a few
discrete field values, is observed for the flakes. Such sharp steps in
magnetization in a ferromagnet are typically observed as Barkhausen jumps
resulting from pinning and depinning of domain walls. We find that these steps
always occur at the same external field values, thereby representing the
magnetization changes due to intrinsic switching of magnetic moments, most
likely in specific layers of FGT-3. Thus, they are expected to occur at certain
specific field values. Based on our observation of the magnetization of the
bulk crystals, it appears that such characteristic switching behavior may be
clearly observed in thinner samples. Thus, to further investigate the detailed
behavior of the switching in thinner samples, we attempted to perform anomalous
Hall effect (AHE) measurements on a few-layer flakes of the FGT-3 crystals. To
avoid any oxidation of surfaces and carry out the magneto-transport experiments
on pristine samples, the entire process of fabrication of the encapsulated
sample of $\sim$\,20 layers was carried out in a glove box as mentioned above.
The level of O$_2$ and humidity of $\leq$\,0.1 ppm is critical to the results
obtained as discussed below.

%%%%%%%%%%%%% fig 

\begin{figure}[h!]
\includegraphics[width=1\linewidth]{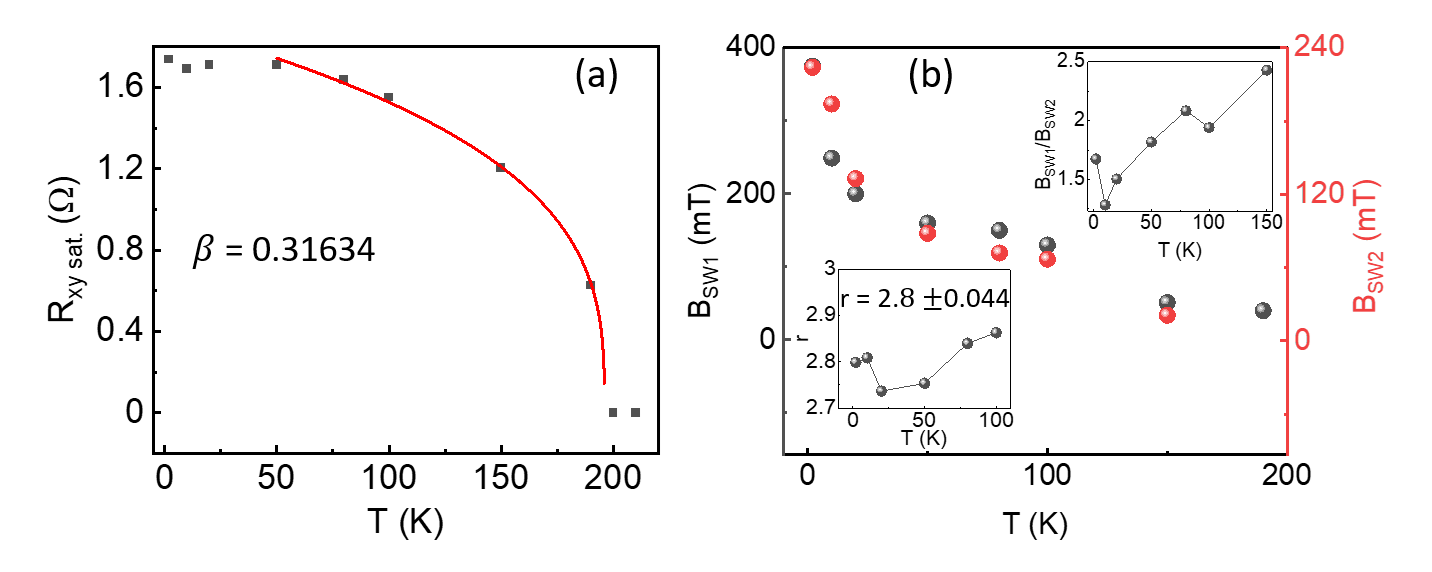}
\caption{(a) Temperature dependence of anomalous Hall 
	resistance ($R_{\rm{xy}}$) at saturation field. The red curve is 
	the critical exponent power law equation fit, where $\beta$ is the 
	critical exponent. (b) $T$-\,dependence of the switching fields at 
	first switching ($B_{\rm{SW1}}$) and second switching ($B_{\rm{SW2}}$). (Inset) 
	$T$-\,dependence of the ratio $B_{\rm{SW1}}/B_{\rm{SW2}}$ (top) and r\,=\,
	$\Delta\,R_{\rm{xy}}\rm{(SW1)}/\Delta\,R_{\rm{xy}}\rm{(SW1)}$ (bottom). 
	Here $\Delta R_\mathrm{{xy}(SW1,2)}$ represent the change in $R_\mathrm{xy}$ at 
        SW1 and SW2, respectively.}
\label{sw field}
\end{figure}
%%%%%%%%%%%%%%%%%%%%%

The thin flake, as shown in Fig.\,\ref{Structural properties}(d), is
characterized by the Raman spectroscopy measurements. The Raman spectrum
exhibits two clear peaks as observed for the bulk sample. An additional peak at
108\,cm$^{-1}$ is also observed for the flake. The intensity of Raman mode near
110\,cm$^{-1}$ becomes more prominent at thin layers, consistent with a
theoretical study by X. Kong, $et$ $al.$ \cite{kong2021thickness}. Thus, we
confirm the pristine crystalline quality of the flake after device fabrication.
Fig.\,\ref{Transport}(b) shows the temperature-dependent resistivity ($\rho
-T$) curve measured for the encapsulated FGT-3 flake. The change in slope,
determined by a derivative of $\rho -T$ curve (see inset), suggests
$T_\mathrm{C}$ of 196\,K, slightly lower than that of the bulk sample,
determined from the magnetization measurements. An upturn in resistivity is
observed below $T\sim$ 20\,K, which is consistent with the data reported in
literature~\cite{zhang2018emergence, huang2022anomalous}. As discussed above,
FGT-3 shows itinerant ferromagnetic behavior due to Fe$^{+3}$ ions and
localized behavior due to Fe$^{+2}$ ions~\cite{sharma2023non}. The presence of
localized magnetic moments effectively acts as internal magnetic impurities,
whose interaction with conduction electrons induces a characteristic Kondo
upturn in resistivity at low temperatures. The AHE measured at $T$\,=\,2\,K
for the device is shown in Fig.\,\ref{Transport}(c). The AHE exhibits two sharp
jumps in $R_{\rm{xy}}$ at external field values of 224\,mT (switching
1\,-\,SW1) and 375\,mT (switching 2\,-\,SW2), respectively while up sweep.
Considering that the AHE is due to the magnetization of the sample, these jumps
suggest two-level switching of magnetization in response to the applied field.
Similar to our observation of field-dependent non-stochastic behavior of the
jumps observed for $M-H$ curves for flakes, these jumps in $R_{\rm{xy}}$ are
also found to occur each time at the same fields during multiple field sweeps
carried out to verify the (non-)stochasticity of the jumps. Moreover, the
change in Hall resistance $\Delta R_{\rm{xy}}$ during the two switching
processes exhibits a characteristic behavior, viz., $\Delta R_{\rm{xy}}$(SW2)
$> $  $\Delta R_{\rm{xy}}$(SW1). This characteristic behavior is consistently
observed in the temperature dependence of $R_{\rm{xy}}$ vs $B$. As shown in
Fig.\ref{Transport}(d), the two sharp switchings in the hysteresis loop of
$R_{\rm{xy}}$ vs $B$ persist till the $T_{\rm{C}}$. However, additional
features such as linear variation of $R_{\rm{xy}}$ vs $B$ after the 1st
switching and a non-linear change before saturation emerge at temperatures
between $T\sim$\,100\,-\,150\,K.

%Fig.\,\ref{Magnetization} (b) also shows a kink feature in the ZFC 
%curve near 160 K at low magnetic fields ($B$) which is absent at a 
%higher magnetic field, $B$\,=\,100\,mT, which agrees well with 
%the earlier reported literature \cite{yi2016competing}.

The ZFC M-T curves (Fig.\,\ref{Magnetization} (b)) show the magnetic moment has
a sudden increase near T $\sim$\,160 K which is due to the dominance of FM
coupling in the system as the AFM coupling vanishes at T $\sim$\,160 K due to
the decrease in the magnetic anisotropy at higher temperatures.  These
observations suggest that the jumps in $R_{\rm{xy}}$ vs $B$ are likely to be
due to an intrinsic magnetization switching process of different inequivalent
Fe atoms in the individual layers. A weak non-linear response of $R_{\rm{xy}}$
is observed till $T\sim$\,200\,K which becomes perfectly linear at
$T$\,=\,210\,K suggesting that the $T_{\rm{C}}$ as observed from the slope
change of $R_{\rm{xx}}$ may not be the true thermodynamic Curie temperature as
indirectly determined from AHE measurements. Such a discrepancy of $T_{\rm{C}}$
from electrical transport and magnetization measurements is often observed as
the resistivity measurements depend on the short-range ordering of magnetic
moments, which may often persist beyond the true thermodynamic value of
$T_{\rm{C}}$, which represents the onset of long-range order in the system. In
such cases, the anomaly in resistivity appears at a higher temperature than the
thermodynamic value of $T_{\rm{C}}$. However, in our case, the observation of
$T_{\rm{C}}$ determined from  $R_{\rm{xy}}$ vs $T$ is lower than that
determined from the response of AHE. This suggests that $T_{\rm{C}}$ determined
from resistivity anomalies may differ from the actual $T_{\rm{C}}$ reported for
thin flakes of such van der Waals materials, where performing magnetization
measurements is challenging due to the small volume of the thin flake samples.
For our FGT-3 samples, the difference (discrepancy!) in $T_{\rm{C}}$ values is
of $\sim$\,5\,K.

From the $T$-dependent data, we note further three important observations - the
$T$-dependence of $R_{\rm{xy}}$ at saturation shows $(T_{\rm{C}}-T)^{\beta}$
type behavior with critical exponent $\beta\,=\,0.316$, See Fig.\ref{sw
field}(a). The value of $\beta=0.316$ suggests a 3D Ising-type interaction in
the system. Secondly, we further observe that the external fields
($B_\mathrm{SW1}$, $B_\mathrm{SW2}$) for the switchings SW1 and SW2,
respectively, systematically reduce as a function of $T$. As fluctuations of
magnetization ($M$) increase with $T$, the uniaxial anisotropy energy density
($K$) reduces with an increase in $T$. Thus, a monotonous behavior of
$B_\mathrm{SW1}$ and $B_\mathrm{SW2}$, as shown in Fig.\,\ref{sw field}(b), are
observed as a function of $T$. Third, the ratio
$B_\mathrm{SW1}$/$B_\mathrm{SW2}$ shows a linearly increasing behavior as a
function of $T$, clearly suggesting that the two sharp switchings are due to
the magnetization reversal of Fe magnetic moments with different
$T$-\,dependent anisotropy behavior. Interestingly, we also find that although
$\Delta R_\mathrm{xy}$ at each switching varies as a fucntion of $T$, the ratio
$ \Delta R_\mathrm{{xy}(SW1)}/\Delta R_\mathrm{{xy}(SW2)}$ remains nearly
constant at $\sim$\,2.8 with slight reduction in the $T$ range of 20\,-\,40\,K,
see left-inset of Fig.\,\ref{sw field}(b).  Here $\Delta
R_\mathrm{{xy}(SW1,2)}$ represent the change in $R_\mathrm{xy}$ at SW1 and SW2,
respectively.

%%%%%%%%%%%%% fig
\begin{figure}[t]
\begin{center}
\includegraphics[scale = 0.35]{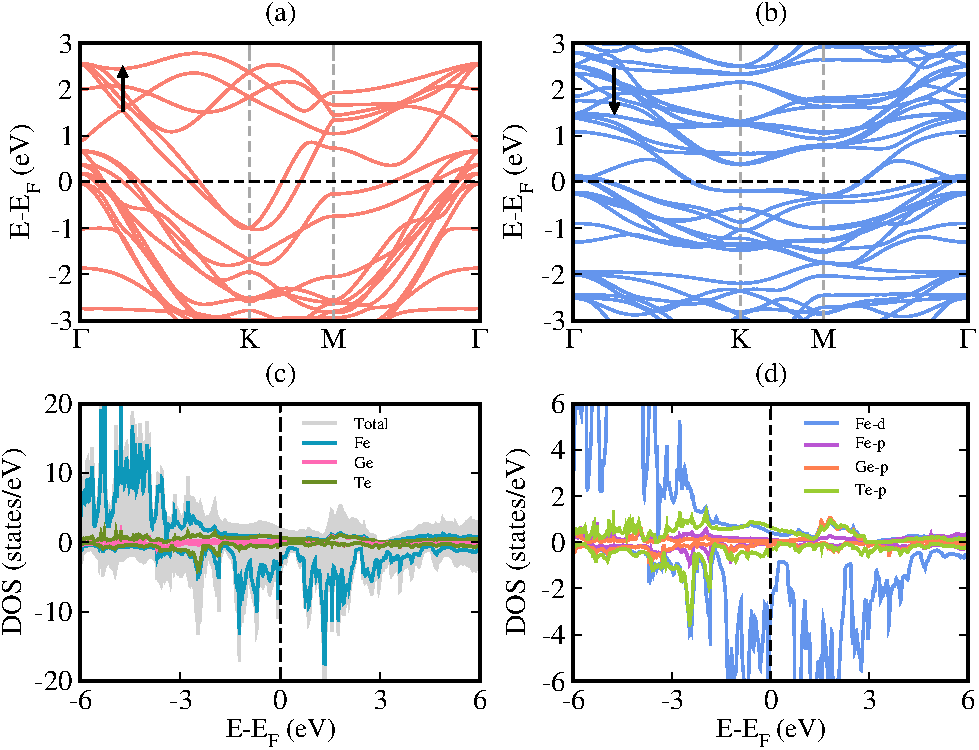}
\caption{The spin polarized electronic band structure of Bulk FGT, 
	(a)\,\&\,(b) spin-up and spin-down channels, respectively. 
	Panels (c)\,\&\,(d) The atom-projected and orbital-projected 
	density of states, respectively. The Fermi level is set to zero.}
\label{band}
\end{center}
\end{figure}
%%%%%%%%%%%%%%%

%%%%%%%%%%%%%%%%% table
\begin{table}[h!]
\caption{The shift in the thermo-hysteresis between FCC \& FCW $M$-$T$ 
	curves towards higher $T$ as the field reduces from 100 mT to 5 mT. 
	Here the irreversibility temperature ($T_\mathrm{ir}$) refers the 
	temperature where the bifurcation between FCC and FCW starts.}
\begin{ruledtabular}
\begin{tabular}{lcr}
%Left\footnote{Note a.}&Centered\footnote{Note b.}&Right\\
%\hline
Irreversibility Temperature $T_\mathrm{ir}$ (K) & $B$ (mT) \\ \hline
130 & 5  \\
120 & 10 \\
    70 & 20 \\
    30 & 100 \\
\end{tabular}
\label{tab:table1}
\end{ruledtabular}
\end{table}
%%%%%%%%%%%%%%%%%%%%%%%%%%%

We note that sharp single switching for thin FGT-3 flakes was reported earlier
for very high-quality devices fabricated in a controlled atmosphere; however,
for devices with top-contact~\cite{fei2018two, deng2018gate,
srivastava2024unusual}. The analysis, as discussed above, suggests that the
sharp two switchings for the FGT-3 device in this case may be related to
coherent magnetization reversal of two different sets of Fe moments in the unit
cells. In order to achieve an in-depth understanding of this double switching
behavior, we carry out a theoretical analysis of magnetization using density
functional theory based first-principles calculations.

From our calculations, we first checked the consistency of our theoretical
analysis with the corresponding reported values in literature. We begin with
the experimental lattice parameters to achieve the ground state of the system.
Our calculated lattice constants (a = b = 3.990 {\AA}, c = 16.329 {\AA}) are
consistent with our experimental reults (a = b = 3.99 {\AA}, c = 16.35 {\AA})
as well as previously reported values (a = b = 3.991 {\AA}, c = 16.33 {\AA}) 
\cite{deiseroth2006fe3gete2}. The optimized structure is further used to 
examine the electronic structure of the material. We have calculated 
spin-polarized band structure and density of states (DOS), as shown in 
Fig. \ref{band}. As discernible from panels (a) and (b) of Fig. \ref{band}, 
FGT-3 exhibits a metallic nature. This is consistent with the metallic nature 
observed for FGT-3~\cite{jiang2022large, kim2022fe3gete2}. In panels (c) 
and (d) of Fig. \ref{band}, we show the total DOS, atom-projected DOS and 
orbital-projected DOS, respectively. Consistent with the bands, the DOS 
also show a metallic behavior, however, with an asymmetry in the states 
associated with spin-up and spin-down channels. The lower valence bands 
in the energy range from -6 to -3 eV mainly originate from the 
hybridization of Fe-$d$, Ge-$p$ and Te-$p$ orbitals. The upper valence and 
lower conduction bands (-2 to 1 eV) are mainly dominated by Fe-$d$ orbitals 
with some minor contributions of Te p-orbitals around the Fermi level. It 
is clear that the metallic character of FGT mainly arises from the $d$-orbital 
of Fe.

%%%%%%%%%%%%%%%%%%%%%%%% table 2
\begin{table}
 \caption{Computed ground state magnetic properties for bulk and monolayer FGT. 
	The bulk results include the total energy (in eV/f.u.) of the FM, AFM configurations and their 
	difference ($\Delta E = E_{\rm FM} - E_{\rm AFM}$) (in meV), magnetic anisotropy energy 
	$\Delta E_{MAE}$ (in meV), magnetic moments of Fe, Ge and Te atoms ($\mu^{\rm Fe^{+3}}$,
	$\mu^{\rm Fe^{+2}}$, $\mu^{\rm Ge}$, $\mu^{\rm Te}$) (in $\mu_B$/atom), total magnetic moment 
	($\mu^{\rm Total}$) (in $\mu_B$/f.u.), and interlayer exchange coupling constant $J_{\rm z}$ (in meV).
	The monolayer results include the total energies for FM and various AFM 
	configurations ($E_{\rm FM}$, $E_{\rm AFM1}$, $E_{\rm AFM2}$, $E_{\rm AFM3}$) (in eV/f.u.), 
	and the exchange coupling constants $J_{\rm 1}$ and $J_{\rm 2}$ (in meV). 
	Reported values are shown for comparison.}
\centering
\begin{ruledtabular}
\begin{tabular}{lrrr}
 &  Our result & Reported result  \\ \hline
 FGT Bulk && \\
$\Delta E_{MAE}$ & 5.63  & 3.37 \cite{jiang2022large}  \\
$E_{\rm FM}$ & -27.1736  & -  \\
$E_{\rm AFM}$ & -27.1660 & -  \\
$\Delta E$ & -15.11  &  -  \\
$\mu^{\rm Fe^{+3}}$  & 2.35  & 2.41 \cite{jiang2022large}  \\
$\mu^{\rm Fe^{+2}}$ &  1.43  & 1.53 \cite{jiang2022large}  \\
$\mu^{\rm Ge}$  &  -0.09  & -0.10 \cite{jiang2022large}  \\
$\mu^{\rm Te}$  & -0.03   &  -0.04 \cite{jiang2022large} \\
$\mu^{\rm Total}$  & 5.97 &  6.29  \cite{jiang2022large} \\
$J_{\rm z}$ & 3.7  &  \\
\hline\
FGT monolayer && \\
$E_{\rm FM}$ & -26.7691    & \\
$E_{\rm AFM1}$ & -26.9105  &  \\
$E_{\rm AFM2}$ & -26.9226  & \\
$E_{\rm AFM3}$ & -26.9514  & \\
$J_{\rm 1}$ & -0.47  & -0.44  \cite{a46}  \\
$J_{\rm 2}$ & 1.24  &  3.27  \cite{a46} \\
$J_{\rm 2}$/$J_{\rm 1}$ & 2.63 &  
\end{tabular}
\end{ruledtabular}
\label{table2}
\end{table}
%%%%%%%%%%%%%%%%%%%%%%%%%%%%%%%%%%%%

To investigate, the magnetic properties of FGT-3 using first-principles 
calculations, we first check the consistency of our results of
computations with the values reported in literature. We first examined the
energetics for both ferromagnetic (FM) and antiferromagnetic (AFM)
configurations for the system. The results of our calculations for total
energies for FM and AFM configurations are given in Table \ref{table2}. The
results show the FM ground state for bulk FGT-3, which has 7.6 meV lower energy
than for the AFM state. This is consistent with our experiments and other
reported literature \cite{deng2018gate,kong2020interlayer}.
We have also computed the magnetic moment of individual elements,
as given in Table \ref{table2}. Our calculated magnetic moments of 
Fe$^{+3}$ \& Fe$^{+2}$ are 2.35 and 1.43 $\mu_B$ per atom, respectively, 
which are in good agreement with the previously reported values 
\cite{jiang2022large}. The dominant contribution to magnetism is 
clearly from the unpaired $d$-electrons of Fe ions.

%%%%%%%%%%%%%%%%% fig
\begin{figure}[h!]
\includegraphics[width=1\linewidth]{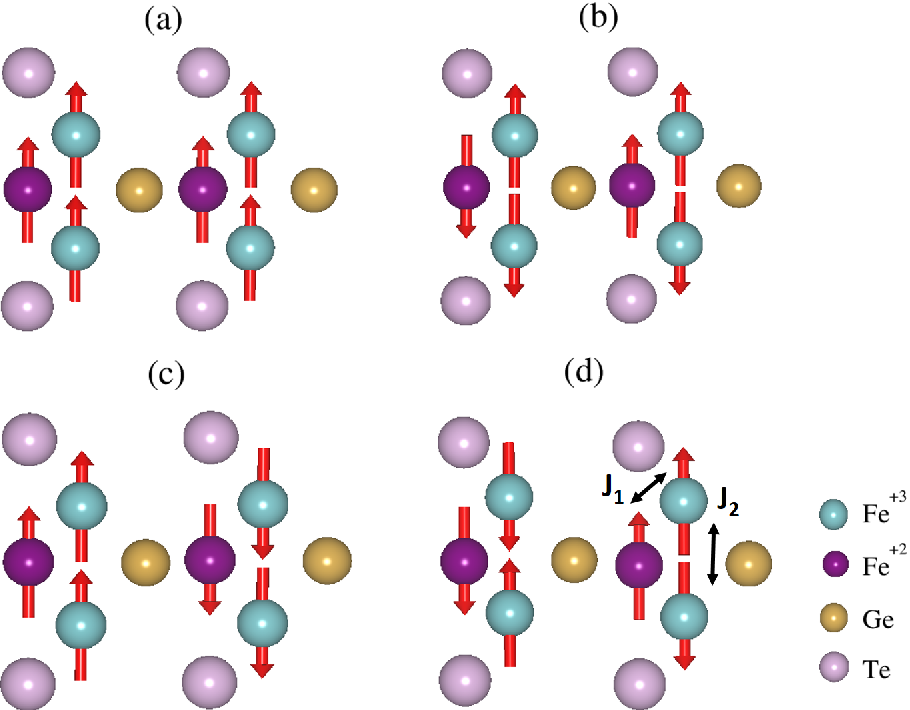}
\caption{The schematic of (a) FM, (b) AFM-1, (c) AFM-2, and (d) AFM-3
        magnetic configurations of monolayer FGT and J$_1$ \& J$_2$ are
        schematic representations of the exchange interaction between Fe
        atoms of the FGT-3 monolayer.}
\label{conf}
\end{figure}
%%%%%%%%%%%%%%%%%%%

Magnetocrystalline anisotropy energy (MAE), defined as $E_{\rm MAE} = E_{\rm
IP} - E_{\rm OOP}$, where $ E_{\rm IP}$ and $E_{\rm OOP}$ is total energy for
in-plane and out-of-plane configurations, respectively, was calculated next. A
positive magnetic anisotropy energy of 5.63 meV was found, indicating an
out-of-plane spin anisotropy for FGT-3, which is also consistent with our
experimental findings and other reported literature \cite{jiang2022large}.

In order to investigate the origin of the sharp double switchings in our AHE
data, we next examined the interlayer and intralayer exchange coupling
constants in FGT-3. Using the Heisenberg model Hamiltonian
%We further calculate the interlayer exchange constant ($J_{\rm{z}}$) = 3.7
%meV, between layers of FGT by using the classical Heisenberg model
%\cite{greiner1995models}.  The classical Heisenberg model is given by:
\[H = E_0 - \sum_{i>j} J_{ij} \left( \hat{e}_i \cdot \hat{e}_j \right)\]

we find interlayer exchange coupling constant ($J_{\rm{z}}$) = 3.7 meV. Here,
$E_0$ is the magnetic configuration independent part of the total energy,
$J_{ij}$ is the exchange coupling parameter, and $\hat{e}_i$ is the unit vector
representing the direction of the magnetic moment on site $i$. The positive
J$_z$ signifies an FM interaction between adjacent layers in FGT-3. 
We next focus on the details of intralayer exchange interactions. 
We examine the magnetic ground state of monolayer FGT-3 using a 
2*1*1 supercell. By fixing a prior spin orientations of Fe$^{+3}$ and 
Fe$^{+2}$ ions in a monolayer, we consider four different possible magnetic 
configurations as shown in Fig.\,\ref{conf}(a)-(d). Depending on the assumed 
spin orientation of Fe atoms, the four magnetic configurations, one FM and 
the other three AFM, viz., AFM1, AFM2, and AFM3, were considered. We
determine the energetics for each of these configurations for monolayer FGT-3,
which we list in Table\,\ref{table2}. From our detailed calculations, we find
the ground state of monolayer FGT-3 is of configuration AFM3 with energy
difference ($\Delta E$) of 182.3\,$\mu$eV, whereas the $\Delta E$ for 
AFM1 configurations is $\sim$141\,$\mu$ eV with respect to the FM 
configuration. Because of such a small energy difference between 
an FM and AFM configurations, very small changes in stoichiometry or 
positional configuration of Fe$^{+3}$ and Fe$^{+2}$ ions may lead 
to stabilization of FM ground state in real crystals. As the unit 
cell of FGT-3 contains Fe$^{+2}$ as well as Fe$^{+3}$ ions, we further 
calculate the exchange coupling constants J$_1$ and J$_2$ between the 
Fe$^{+3}$ - Fe$^{+3}$ and Fe$^{+2}$\,-\,Fe$^{+3}$ ions, respectively for 
the ground state configuration, see Fig.\,\ref{conf} (d). We find J$_1$\,=\,1.24\,meV 
and J$_2$\,=\,-0.47\,meV. Thus, these calculations suggest an AFM interaction 
between Fe$^{+3}$\,-\,Fe$^{+3}$ spins whereas a FM interaction is found between 
Fe$^{+3}$ and Fe$^{+2}$ spins.

%%%%%%%%%%%% fig

\begin{figure}[h!]
\includegraphics[width=1\linewidth]{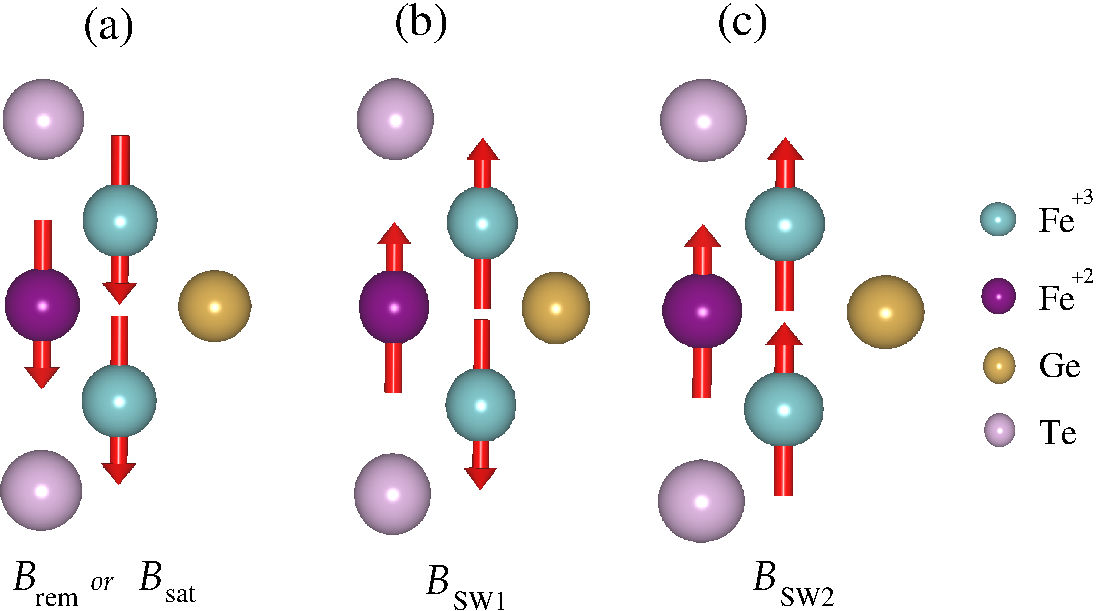}
\caption{The schematic of magnetic moments of Fe$^{+3}$
        and Fe$^{+2}$ at remanence ($B_{\rm{rem}}$) or saturation ($B_{\rm{sat}}$) and two
        different switching fields ($B_{\rm{SW1}}$) \& ($B_{\rm{SW2}}$).}
\label{Figmodel}
\end{figure}

%%%%%%%%%%%%%%%%%

In order to validate the two types of exchange interactions among the different
Fe-ions, we next analyze the changes in the magnetization ($\Delta M_1$ and
$\Delta M_2$) during the switchings at external fields $B_{\rm{SW1}}$ and
$B_{\rm{SW2}}$, respectively. Fig.\,\ref{Figmodel} shows the behavior of
Fe$^{+3}$ and Fe$^{+2}$ moments at remanence and two different switching
fields. The remanence and saturation states are the same. While upsweeping the field from remanent state, as the field is
increased in the opposite direction, all ferromagnetically coupled Fe$^{+3}$
and Fe$^{+2}$ moments switch at $B_{\rm{SW1}}$, thereby inducing a sharp
step-like change in magnetization $\Delta M_1$ and thereby the corresponding
AHE $\Delta R_{xy}$(SW1). The AFM coupled Fe$^{+3}$ spins switch at higher
field ($B_{\rm{SW2}}$). The change in magnetization $\Delta M_2$ due to this
switching is less than for the 1st switching, as in this case, only those spins
that are oppositely directed switch. The corresponding change in $\Delta
R_{xy}$(SW2) as observed in our experiment is about $\frac{1}{3}$rd of $\Delta
R_{xy}(SW1)$. Thus, we find a qualitative agreement between our observed
experimental results of $\Delta R_{xy}$(SW1) and $\Delta R_{xy}$(SW2) and
theoretical calculations. Furthermore, we also observe that the ratio
$\rm{J_1}/\rm{J_2} \approx 2.63$, which is nearly of similar value as that for
the ratio $\Delta R_{xy}(\rm{SW1})/\Delta R_{xy}(\rm{SW2})$ obtained from our
experimental measurements. As discussed above, $\Delta R_{xy}$(SW1) and $\Delta
R_{xy}$(SW2) correspond to switchings of ferromagnetically ($\sim \rm{J}_1$)
and antiferromagnetically aligned spins ($\sim \rm{J}_2$), respectively.

Finally, we note from the temperature dependent $R_{xy}$ vs $B$ data that the
2nd switching vanishes between $T\sim$ 150\,K and 190\,K which is again
consistent with the observation of a kink-like feature at $T\sim$ 160\,K
observed in the $M-T$ (ZFC) data as shown in Fig.\,\ref{Magnetization}(b). As
discussed above, the kink-like feature is most likely due to the AFM transition
($T_N$) at $\sim$160\,K. The observed $T_\mathrm{N} < T_\mathrm{C}$ is also consistent with the
calculated values of corresponding exchange coupling values, viz., $J_1 < J_2$.

Thus, from all our bulk magnetization and AHE results as well as first-
principles calculations, we find a consistent and clear signature of an
intra-layer antiferromagnetic coupling between Fe$^{+3}$ and Fe$^{+3}$ moments
in FGT-3 crystals. Our results show that detailed analysis of bulk
magnetization and magnetotransport behavior of high-quality samples fabricated
in a very clean environment is important to extract intrinsic behavior of
magnetization of such van der Waals magnets. Our work may pave the way to
exploring new magnetic phases and designing future spintronic devices.

%%%%%%%%%%%%%%%%%%%%%%%%%
\begin{acknowledgments}
PD gratefully acknowledges the funding support from the Science \& Engineering
Research Board (SERB) of Govt. of India (grant no. SPR/2021/000762). BKM acknowledges the funding support from SERB, DST (CRG/2022/000178). NY is
thankful to the Council of Scientific \& Industrial Research (CSIR) for
financial support. Shivani acknowledges the fellowship support from UGC (BININ01949131),
Govt. of India. Sandeep is thankful to Ministry of Education (MoE) for the financial support. NY, Sandeep \& PD acknowledge the Dept of Physics, Central Research
Facility (CRF) and Nano Research Facility (NRF), Indian Institute of Technology
(IIT) Delhi, for availing of various experimental facilities. Shivani \& BKM are thankful to High 
Performance Computing cluster Tejas at the Indian Institute of 
Technology Delhi and PARAM Rudra, a national supercomputing
facility at Inter-University Accelerator Centre (IUAC), New Delhi.

\end{acknowledgments}
%%%%%%%%%%%%%%%%%%%%%%%%%%

\section*{Data Availability Statement}

The data supporting this study's findings are available from the corresponding
author upon reasonable request.

\nocite{*}

%%%%%%%%%%%%%%%%%%%%%%%%%%%%%%%%%%%%%%%%%%%

%%%%%%%%%%%%%%%%
%\bibliographystyle{apsrev4-2}
\bibliography{FGT}

\end{document}